Research article

# Study of hadronic event shape in flavour tagged events in $e^+e^-$ annihilation at $\langle\sqrt{s}\rangle$ = 197 GeV

L3 Collaboration

Address: CERN, Geneva, Switzerland

Email: L3 Collaboration - Salvatore.Mele@cern.ch

## Abstract

Results are presented from a study of the structure of hadronic events in high-energy $e^+e^-$ interactions detected by the L3 detector at LEP. Various event shape distributions and their moments are measured at several energy points at and above the Z-boson mass. The event flavour is tagged by using the decay characteristics of b-hadrons. Measurements of distributions of event shape variables for all hadronic events, for light (u, d, s, c) and heavy (b) quark flavours are compared to several QCD models with improved leading log approximation: JETSET, HERWIG and ARIADNE. A good description of the data is provided by the models.

## 1 Introduction

Hadronic events produced in $e^+e^-$ annihilation have been a powerful tool to test the predictions of Quantum Chromodynamics (QCD) [1-5]. Perturbative QCD successfully accounts for many aspects of the hadronic decays of the Z boson [6]. The primary quarks from Z-boson decays first radiate gluons, which in turn may split into quark or gluon pairs. The quark and gluons then fragment into observable hadrons. Perturbative QCD itself does not describe the fragmentation process. Instead several phenomenological models have been developed to describe fragmentation. These models provide a way to correct for the effects of fragmentation in the experimental data, which can then be compared with the perturbative QCD calculations directly.

The event shape variables which characterize the global structure of hadronic events are among the simplest experimental measurements sensitive to the parameters of perturbative QCD and fragmentation models. This article reports on the measurement of event shapes for hadronic events collected at LEP by the L3 detector [7-10] at $e^+e^-$ centre-of-mass energies $\sqrt{s} \geq 189$ GeV. Similar analyses were reported by all LEP experiments [11-15].

Heavy flavour production in $e^+e^-$ annihilation can be studied by exploiting the characteristics of heavy flavour decays. In the present study, hadronic events are separated into heavy (b) and light (u, d, s, c) flavours, and event shape variables are separately measured for these final states. This allows to test the modelling of heavy flavour mass effects. Earlier and similar measurements, at lower centre-of-mass energies, are reported in References [11] and [16].

## 2 Global event shape variables

Event shape variables, insensitive to soft and collinear radiation, are built from linear sums of measured particle momenta. They are sensitive to the amount of hard-gluon radiation. Six global event shape variables are measured here, using calorimetric and tracking information measured as described in References [7-10] and [11]. They are: thrust, scaled heavy jet mass, total and wide jet broadening, the *C*-parameter and the jet resolution parameter. These event-shape variables are defined below.

### 2.0.1 Thrust

The global event-shape variable thrust, *T*, [17,18] is defined as

$$T = \frac{\Sigma|\vec{p}_i \cdot \vec{n}_T|}{\Sigma|\vec{p}_i|}$$

where $\vec{p}_i$ is the momentum vector of particle *i*. The thrust axis $\vec{n}_T$ is the unit vector which maximizes the above expression. The value of the thrust can vary between 0.5 and 1.0. The plane normal to $\vec{n}_T$ divides space into two hemispheres, $S_\pm$, which are used in the following definitions.

### 2.0.2 Scaled heavy jet mass

The heavy jet mass, $M_H$, is defined [19-21] as

$$M_H = \max\,[M_+, M_-],$$

where $M_\pm$ are the invariant masses in the two hemispheres, $S_\pm$,

$$M_\pm^2 = \left[\sum_{i \in S_\pm} p_i\right]^2$$

where $p_i$ is the four-momentum of particle *i*. The scaled heavy jet mass, $\rho_H$, is defined as

$$\rho_{\mathrm{H}} = \frac{M_{\mathrm{H}}^2}{s}.$$

#### 2.0.3 Jet broadening variables

These variables are defined [22,23] by computing in each hemisphere the quantity

$$B_{\pm} = \frac{\sum_{i \in S_{\pm}} |\vec{p}_i \times \vec{n}_T|}{2\sum_i |\vec{p}_i|}$$

in terms of which the total jet broadening, $B_{\mathrm{T}}$, and wide jet broadening, $B_{\mathrm{W}}$, are defined as

$$B_{\mathrm{T}} = B_+ + B_- \text{ and } B_{\mathrm{W}} = \max(B_+, B_-). \quad (1)$$

#### 2.0.4 C-parameter

The $C$-parameter is derived from the eigenvalues of the linearized momentum tensor [24,25]:

$$\Theta_{ij} = \frac{\sum_a p_a^i p_a^j / |\vec{p}_a|}{\sum_a |\vec{p}_a|} \quad i, j = 1, 2, 3;$$

where $a$ runs over final state hadrons and $i, j$ indicate components of the momentum vectors $\vec{p}_a$. With $\lambda_1$, $\lambda_2$ and $\lambda_3$ the eigenvalues of $\Theta$, the $C$-parameter is defined as

$$C = 3(\lambda_1\lambda_2 + \lambda_2\lambda_3 + \lambda_3\lambda_1).$$

#### 2.0.5 Jet resolution parameter

Jets are reconstructed using the JADE algorithm [26,27]. The value of the "closeness variable" at which the classification of an event changes from 2-jet to 3-jet is called the 3-jet resolution parameter $y_{23}^{\mathrm{J}}$.

## 3 Monte Carlo models

The measured global event shape variables are compared below with the predictions of three Monte Carlo parton shower models JETSET[28], ARIADNE[29] and HERWIG[30-32]. In these models parton showers are generated perturbatively according to a recursive algorithm down to energy scales of 1–2 GeV defining a boundary between perturbative and non-perturbative regions of phase space. In the non-perturbative region, hadrons are generated according to phenomenological fragmentation models. In the perturbative phase of all the models, the parton branching energy fractions are distributed according to the leading order DGLAP [33-36] splitting functions.

The basic Leading Logarithmic Approximation (LLA) [37-41] of the models is modified, in the framework of the Modified Leading Logarithmic Approximation (MLLA) [42-44], to take into account certain interference effects first occurring in the Next-to-Leading Logarithmic Approximation (NLLA) [45-48].

The JETSET parton shower Monte Carlo program uses, as evolution variable in the parton shower, the mass squared of the (time-like virtual) branching parton. Angular ordering to describe NLLA interference effects is implemented in an *ad hoc* manner and the distributions of the first generated gluon are reweighted to match those of the tree-level O($\alpha_s$) matrix element. Partons are hadronized according to a string fragmentation model. For light quarks (u, d, s) the Lund symmetric fragmentation function [49] is used and for b and c quarks the Peterson fragmentation function [50]. The transverse momenta of hadrons are described by Gaussian functions.

The parton cascade of ARIADNE evolves via two-parton colour-dipole systems. Gluon radiation splits a primary dipole into two independent dipoles, the evolution variable being the square of the transverse momentum of the radiated gluon. This procedure incorporates, to MLLA accuracy, the NLLA interference effects that give angular ordering in the parton shower. Hadrons are generated according to the same string fragmentation model as used in JETSET.

The HERWIG Monte Carlo program uses a coherent parton branching algorithm with phase space restricted to an angle-ordered region. The evolution variable is $E^2(1 - \cos\theta)$ where $E$ is the energy of the initial parton and $\theta$ the angle between the branching partons. This choice incorporates NLLA interference effects within the MLLA framework. As in JETSET the distributions of the most energetic gluon are improved by matching them to those given by the $O(\alpha_s)$ matrix element. Hadronization is described by a cluster model based on perturbative-level QCD pre-confinement.

The parameters of the models, which are detailed in Reference [11], are tuned, using Z-peak data, by fitting the models to the following distributions:

jet resolution parameter $y_{23}^{\mathrm{J}}$ of the JADE algorithm [26,27];

Fox-Wolfram moment $H_4$ [51-53];

narrow-side minor $T_{minor}^{NS}$ [54];

charged particle multiplicity $N_{\mathrm{ch}}$.

**Table 1: Summary of integrated luminosity and number of selected hadronic events at the different energies.**

| $\sqrt{s}$ (GeV) | Integrated Luminosity ($pb^{-1}$) | Selection Efficiency (%) | Sample Purity (%) | Selected events |
|---|---|---|---|---|
| 188.6 | 175.1 | 87.72 ± 0.62 | 80.92 ± 0.25 | 4473 |
| 191.6 | 29.4 | 87.77 ± 0.62 | 80.11 ± 0.26 | 720 |
| 195.5 | 83.4 | 88.41 ± 0.63 | 78.60 ± 0.27 | 1884 |
| 199.5 | 81.2 | 88.51 ± 0.62 | 77.54 ± 0.25 | 1835 |
| 201.7 | 36.5 | 89.02 ± 0.63 | 76.98 ± 0.25 | 817 |
| 205.1 | 70.5 | 88.77 ± 0.64 | 75.65 ± 0.22 | 1496 |
| 206.5 | 126.2 | 88.93 ± 0.63 | 75.26 ± 0.22 | 2688 |
| 197.0 | 602.2 | 88.33 ± 0.28 | 78.19 ± 0.11 | 13913 |

The last line corresponds to the average $\langle \sqrt{s} \rangle$ and the total sample.

The variable $y_{23}^{J}$ is particularly sensitive to the 3-jet rate, $H_4$ to the inter-jet angles, $T_{minor}^{NS}$ to the lateral size of quark jets and so to the transverse momentum distribution of hadrons relative to a jet axis, and $N_{ch}$ to parameters of the fragmentation models. The tuning was performed independently for all and udsc quark flavours.

More details on the Monte Carlo models and the tuning procedure can be found in Reference [11].

## 4 Data and Monte Carlo samples

The data discussed in this analysis correspond to an integrated luminosity of 602.2 $pb^{-1}$, collected during the years 1998–2000 at $\sqrt{s}$ ≥ 189–207 GeV as detailed in Table 1. Only data corresponding to data-taking periods where all sub-detectors were fully operational are retained in this analysis.

The primary trigger for hadronic events requires a total energy greater than 15 GeV in the calorimeters. This trigger is in logical OR with a trigger using the barrel scintillation counters and with a charged-track trigger. The combined trigger efficiency for the selected hadronic events exceeds 99.9%.

The selection of $e^+e^- \to q\bar{q} \to$ *hadrons* events is based on the energy measured in the electromagnetic and hadron calorimeters, as described in Section 3 of Reference [11]. Energy clusters in the calorimeters are selected with a minimum energy of 100 MeV. The principal variables used to distinguish these hadronic events from background are the cluster multiplicity and energy imbalances. Energy clusters in the calorimeters are used to measure the total visible energy $E_{vis}$, and the energy imbalances parallel and perpendicular to the beam direction:

$$E_{\|} = \left| \sum_i E_i \cos\theta_i \right|,$$

$$E_{\perp} = \sqrt{\left( \sum_i E_i \sin\theta_i \sin\phi_i \right)^2 + \left( \sum_i E_i \sin\theta_i \cos\phi_i \right)^2},$$

respectively, where $E_i$ is the energy of cluster $i$ and $\theta_i$ and $\varphi_i$ are its polar and azimuthal angles with respect to the beam direction.

Monte Carlo events are used to estimate the efficiency of the selection criteria and purity of the data sample. Monte Carlo events for the process $e^+e^- \to q\bar{q} \to$ *hadrons* are generated by the parton shower programs PYTHIA[55] for $\sqrt{s}$ = 189 GeV and KK2F[56,57], which uses PYTHIA for hadronization, for the highest energies. QCD parton shower and fragmentation process are taken from JETSET 7.4 [28]. The generated events are passed through the L3 detector simulation [58,59]. The background events are simulated with PYTHIA and PHOJET[60,61] for hadron production in two-photon interactions, KORALZ[62] for $\tau^+\tau^-$ final state, BHAGENE[63,64] for Bhabha events, KORALW[65,66] for W-boson pair-production and PYTHIA for Z-boson pair-production.

## 5 Event selection and flavour tagging

This analysis has two main sources of background. The first is the so called "radiative return" events, where initial state radiation results in a mass of the hadronic system close to the Z boson. The second is pair-production of W or Z bosons where one or both of the bosons decay hadronically. Additional background arises from hadron production in two-photon interactions and $\tau$ pair production. Events are first selected by requiring $E_{vis}/\sqrt{s} > 0.7$, $E_{\perp}/E_{vis} < 0.4$, number of clusters > 12, and at least one well-measured charged track. To reduce the radiative return background, events are rejected if they have a high-energy photon candidate, defined as a cluster in the electromagnetic calorimeter with at least 85% of its energy in a 15° cone and a total energy greater than $0.18\sqrt{s}$. Radiative return events, where an unobserved photon is emitted close to the beam axis, are reduced by requiring $\sqrt{s'/s} > 0.85$, where $\sqrt{s'}$ is given by

$$\sqrt{s'/s} = \sqrt{1 - 2 \cdot \frac{E_\gamma}{\sqrt{s}}}$$

and the energy of the unobserved photon $E_\gamma$ is derived by first forcing the event into a two-jet topology and then using the angles of the two jets, $\theta_1$ and $\theta_2$, as:

$$E_\gamma = \sqrt{s} \cdot \frac{|\sin(\theta_1+\theta_2)|}{\sin\theta_1+\sin\theta_2+|\sin(\theta_1+\theta_2)|}$$

To reject boson pair-production events where one of the bosons decays into leptons, events having an electron or muon with energy greater than 40 GeV are removed. Hadronic decays of boson pair events are rejected by:

1. forcing the event to a 4-jet topology using the Durham jet algorithm [67-70],

2. performing a kinematic fit imposing energy-momentum conservation,

3. applying cuts on the energies of the most- and the least-energetic jets and on the jet resolution parameter, $y_{34}^{\mathrm{D}}$ at which the event classification changes from 3-jet to 4-jet. Events are rejected if the energy of the most energetic jet is less than $0.4\sqrt{s}$, the ratio of the energy of the most energetic jet to the least energetic jet is less than 5, $y_{34}^{\mathrm{D}} > 0.007$, there are more than 40 clusters and more than 15 charged tracks, and $E_{||} < 0.2E_{\mathrm{vis}}$ after the kinematic fit.

This selection removes 11.67 ± 0.28% of the signal events, 98.11 ± 0.02% of the radiative return events, 83.31 ± 0.03% and 80.08 ± 0.11%, respectively, of W-boson and Z-boson pair-production events. We select a total of 13913 hadronic events, with an efficiency of 88.33 ± 0.28% and with a purity of 78.19 ± 0.11%. The backgrounds due to radiative return, W-boson pairs, Z-boson pairs and hadron production in two-photon interaction are 5.71 ± 0.06%, 12.28 ± 0.04%, 1.01 ± 0.01% and 2.55 ± 0.09%, respectively. The remaining backgrounds are negligible. The integrated luminosity and the number of selected events for each energy point are summarized in Table 1.

Heavy (b) flavour events are separated from light (u, d, s, c) flavour events by using the characteristic decay properties of the b-hadrons. As the first step, the interaction vertex is estimated fill-by-fill by iteratively fitting all the good tracks measured in the detector during the fill. Measurements of all $n$ tracks in the event contribute to a probability, $P^{[n]}$, that all tracks in the event originate from the interaction vertex. This probability is flat for zero lifetime of all produced particles but otherwise peaks at zero. A weighted discriminant is used:

$B_{\mathrm{n}} = -\log P$, where $P = P^{[n]}\sum_{j=0}^{n-1}(-\log P^{[n]})^j / j!$ and $P^{[n]} = \prod_{j=1}^{n} P_j$ and $P_j$ is the probability that track $j$ originates from the primary vertex [71].

Figure 1 shows the distribution of the discriminant $B_n$ for data as well as expectations from signal and background. A cut on this discriminant is made to distinguish events with b-quarks from events without. These two samples are called "b-events" and "non-b events" in the following. The non-b events are selected using $B_n < 1.0$. The b-events are selected with a cut on $B_n > 3.4$. A total of 440 b-events are selected with an efficiency of 26.2 ± 0.4% and a purity of 75.2 ± 1.2% while 6895 non-b events are selected with a selection efficiency of 75.5 ± 0.3% and a purity of 72.7 ± 0.1%. The dominant background for the b-events are due to wrong flavour events amounting to 14.3 ± 0.5% while that due to ISR, W-boson and Z-boson pair events are respectively 4.5 ± 0.3%, 4.5 ± 0.1% and 1.4 ± 0.1%. On the other hand, the dominant background for non-b events are from W-boson pair events amounting to 17.6 ± 0.1% while those due to wrong flavour type, ISR, Z-boson pair and 2-photon events are 3.9 ± 0.1%, 3.7 ± 0.1%, 0.6 ± 0.1% and 1.4 ± 0.1% respectively.

## 6 Measurements

The distributions of event shape variables are measured at each energy point listed in Table 1. The data distributions are compared to a sum of the signal and the different background Monte Carlo distributions obtained using the same selection procedure and normalized to the integrated luminosity according to the Standard Model cross sections. Figures 2 and 3 show the measured distributions for event thrust and total jet broadening for all data, b-events and non-b events.

Data at the different energy points are combined at the average centre-of-mass energy $\langle \sqrt{s} \rangle$ = 197 GeV. The distributions are compared to predictions from signal and background Monte Carlo programs. There is generally good agreement between data and Monte Carlo particularly for the entire sample thus justifying the use of the latter to obtain the correction from detector level to particle level. For Monte Carlo events, these event shape variables are calculated before (particle level) and after (detector level) detector simulation. The calculation before detector simulation takes into account all stable charged and neutral particles. The measured distributions at detector level differ from the ones at particle level because of detector effects, limited acceptance and finite resolution.

After subtracting the background events the measured distributions are corrected for detector effects, acceptance and resolution, on a bin-by-bin basis by comparing the detector level results with the particle level results. In the extraction of flavour-tagged distributions, the contribution of wrong-flavour contamination is subtracted in the same way as the SM background subtraction.

The data are corrected for initial and final state photon radiation bin-by-bin using Monte Carlo distributions at particle level with and without radiation. The comparison between data and Monte Carlo models shown in Figures 4, 5, 6, 7, 8, 9 below is made for particle level distributions.

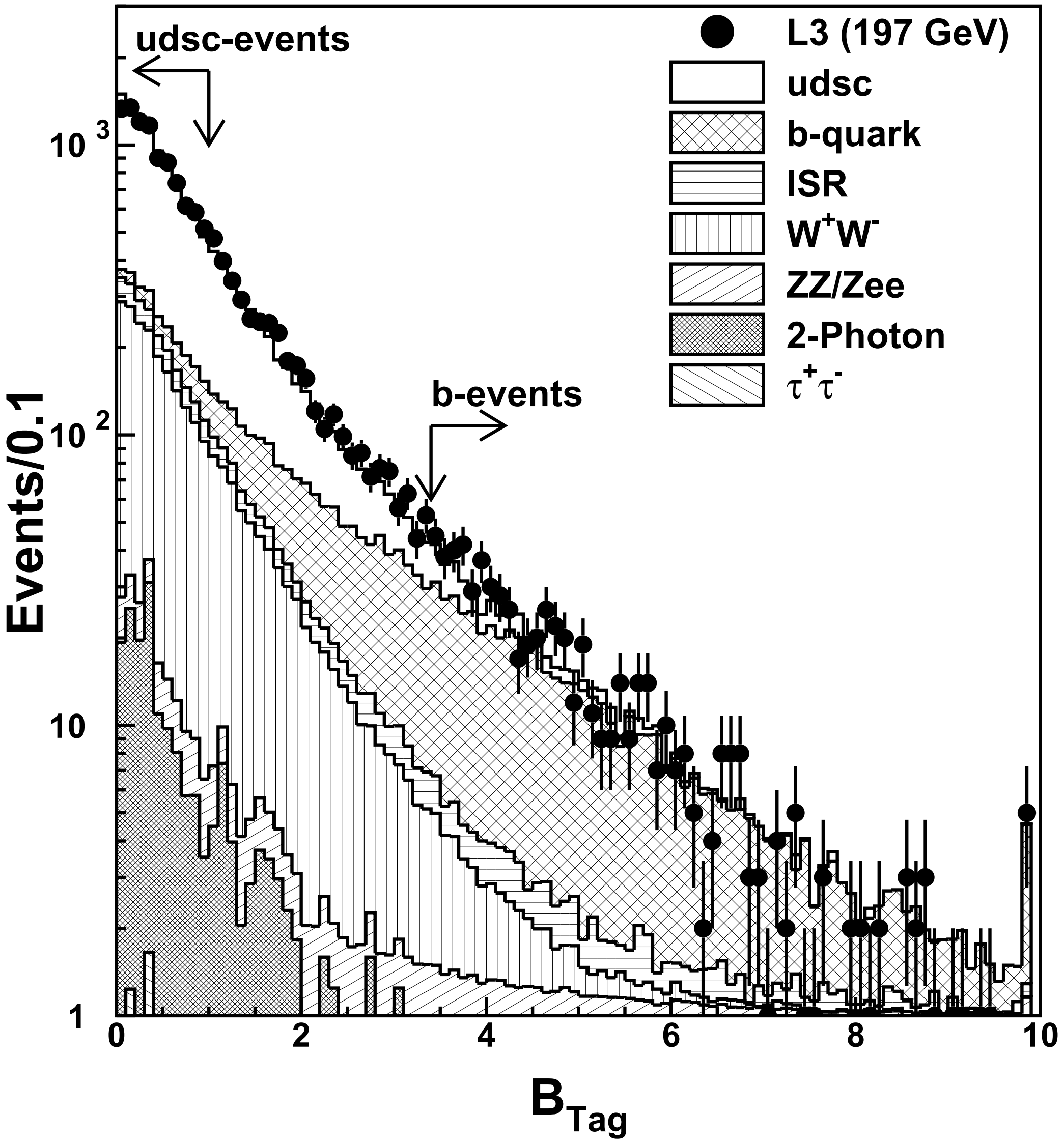

**Figure 1**
**Distribution of the flavour tagging discriminator $B_n$ for the combined data sample together with expectations from signal and background.** The non-b events are selected using $B_n < 1.0$. The b-events are selected with a cut on $B_n > 3.4$.

## 7 Systematic uncertainties

The systematic uncertainties in the distributions of event shape variables are calculated for each bin of these distributions. The main sources of systematic are uncertainties in the estimation of the detector corrections and the background levels.

The uncertainty in from detector corrections is estimated by repeating the measurements altering several independent aspects of the event reconstruction, and taking the largest variation with respect to the original measurement. These changes are:

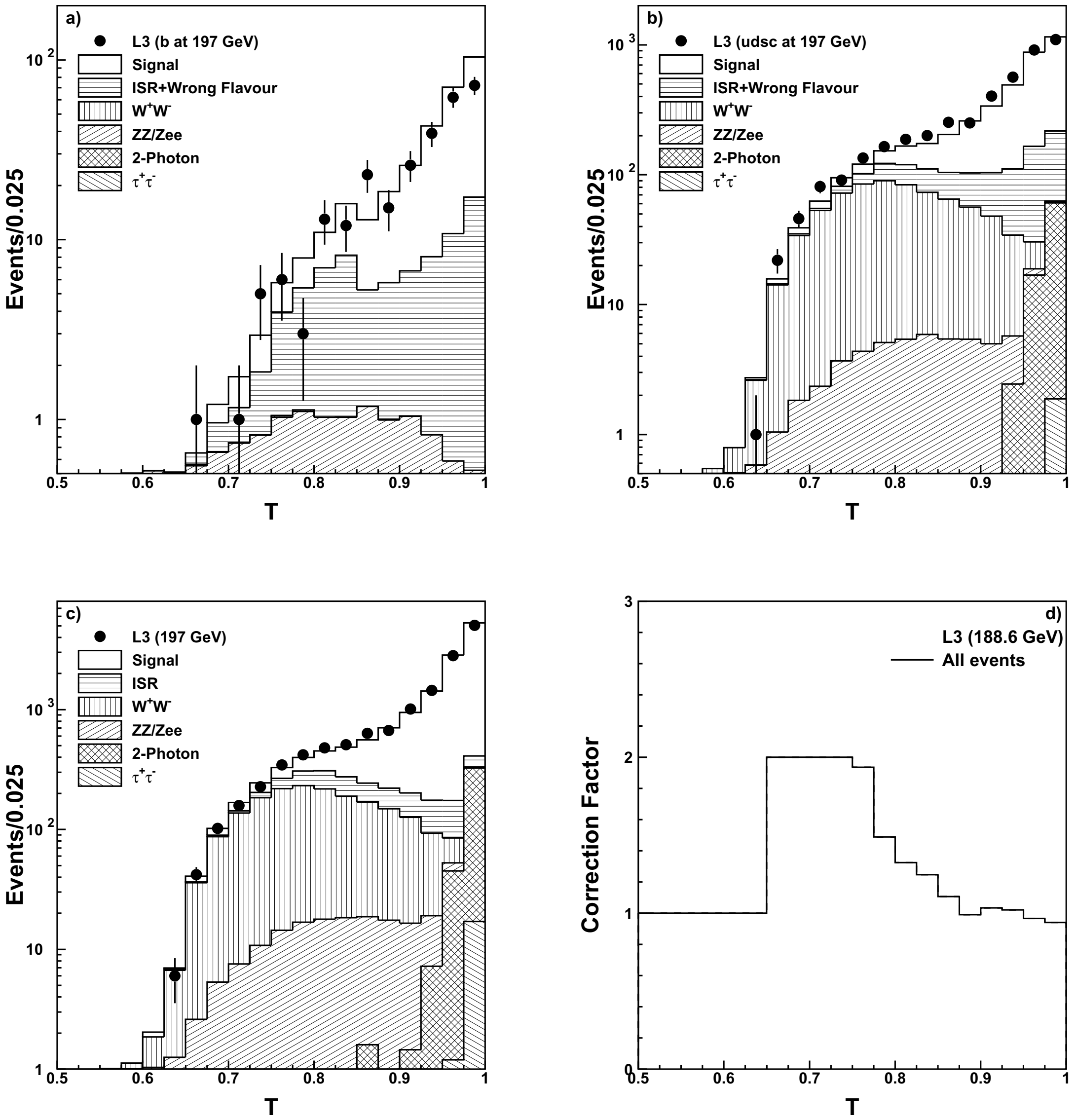

**Figure 2**

**Thrust distribution at detector level at $\langle \sqrt{s} \rangle$ = 197 GeV measured for (a) b-events (b) non-b events and (c) all events.** The solid lines correspond to the overall expectation from theory. The shaded areas refer to different backgrounds and the white area refers to the signal as predicted by PYTHIA and KK2F. The correction factor to pass from the observed distributions, after background subtraction, to the measured event-shape variable is presented in (d) for the inclusive sample without flavour tag for a centre-of-mass energy $\sqrt{s}$ = 188.6 GeV.

• the definition of reconstructed objects used to calculate the observables is changed from calorimetric clusters only to a non-linear combination of charged tracks with calorimetric clusters;

• the effect of different particle densities in correcting the measured distributions is estimated by using a different signal Monte Carlo program, HERWIG instead of JETSET or PYTHIA;

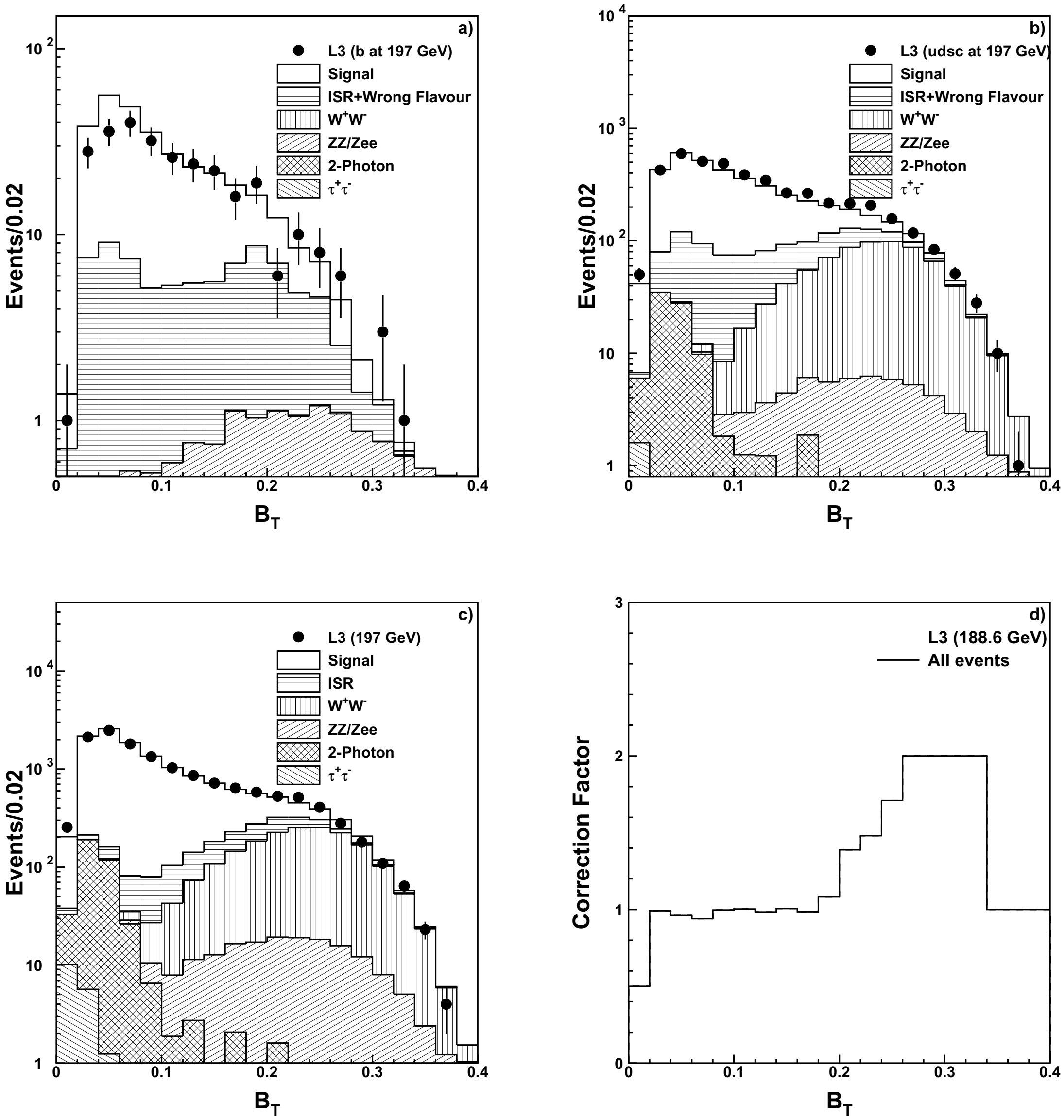

**Figure 3**

**Measured total jet broadening distribution at $\langle \sqrt{s} \rangle$ = 197 GeV for for (a) b-events (b) non-b events and (c) all events.** The solid lines correspond to the overall expectation from theory. The shaded areas refer to different backgrounds and the white area refers to the signal as predicted by PYTHIA and KK2F. The correction factor to pass from the observed distributions, after background subtraction, to the measured event-shape variable is presented in (d) for the inclusive sample without flavour tag for a centre-of-mass energy $\sqrt{s}$ = 188.6 GeV.

• the acceptance is reduced by restricting the events to the central part of the detector, $|\cos(\theta_T)| < 0.7$, where $\theta_T$ is the polar angle of the thrust axis relative to the beam axis.

The systematic uncertainties on the background levels are assessed by varying the procedure used for the background evaluations and taking the the difference with the original measurements. These changes are:

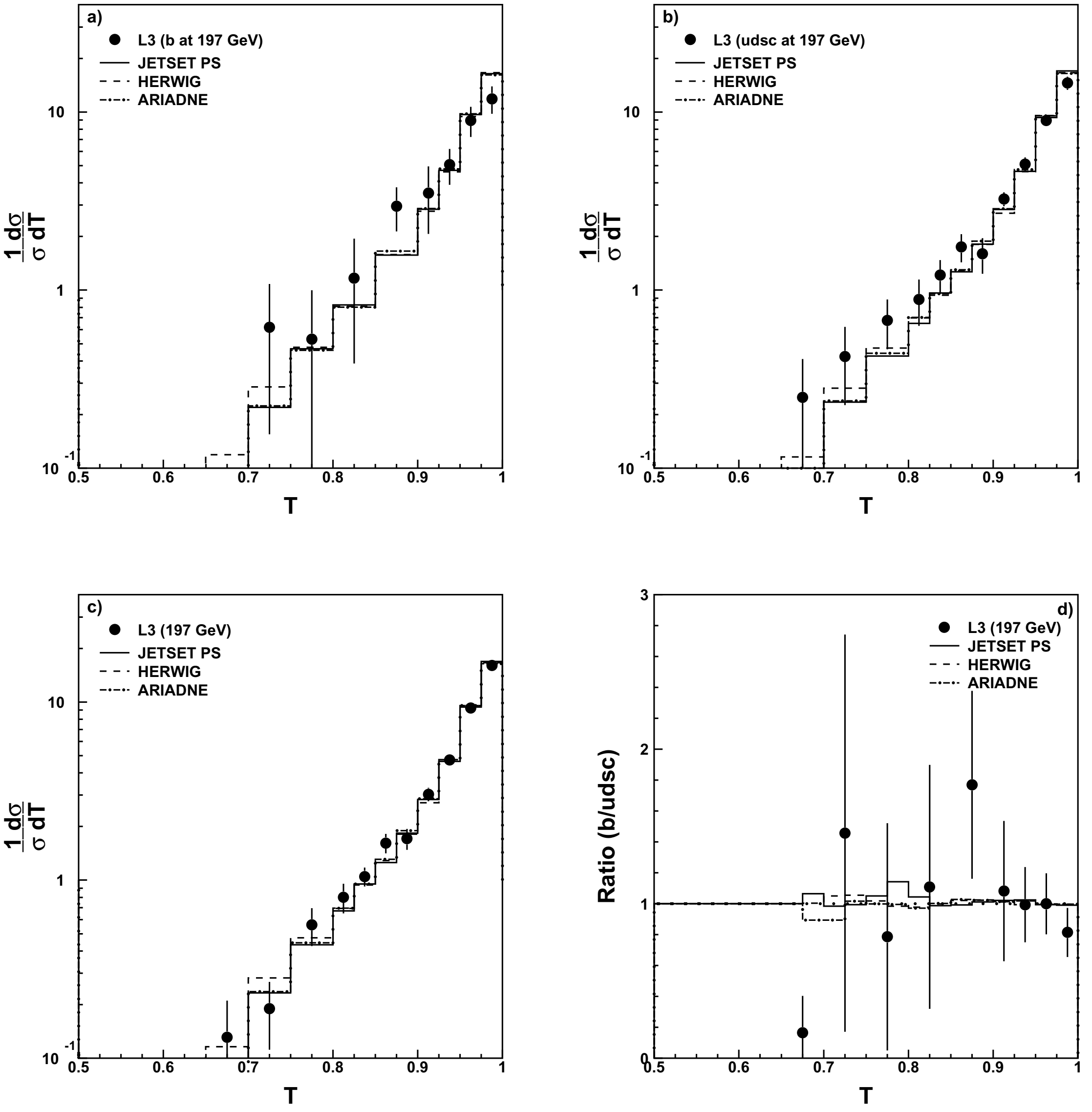

**Figure 4**

**Thrust distributions at $\langle\sqrt{s}\rangle$ = 197 GeV for a) b-events, b) non-b events, c) all events and d) the ratio between b- and non-b events compared to several QCD models.** The error bars include both statistical and systematic uncertainties.

• an alternative criterion is applied to reject radiative return events based on a cut in the two dimensional plane of $E_{||}/E_{vis}$ and $E_{vis}/\sqrt{s}$ ;

• the estimated background from two-photon interaction is varied by ± 30% and is simulated by using the PHOJET instead of the PYTHIA Monte Carlo program;

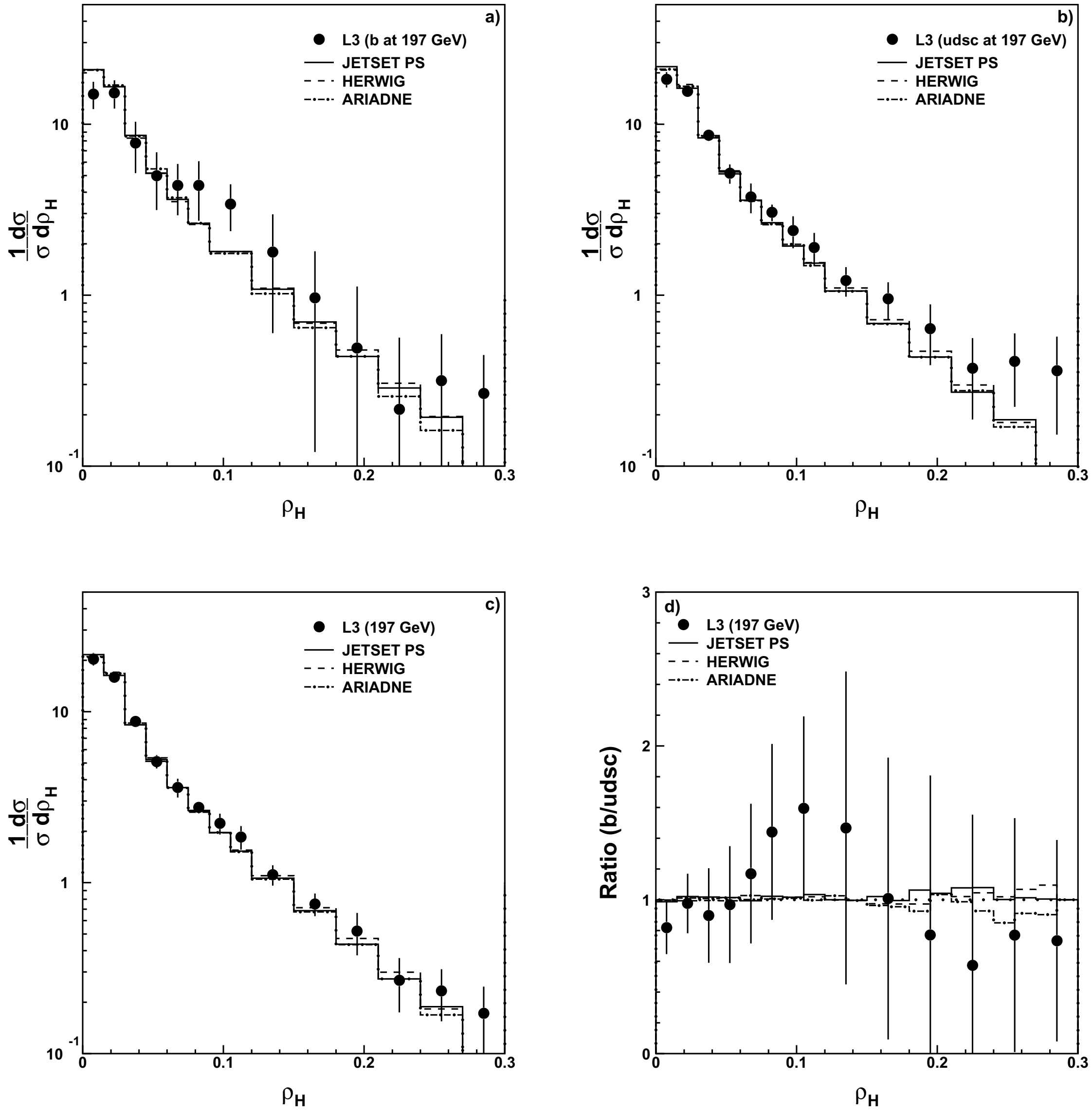

**Figure 5**

**Scaled heavy jet mass distributions at $\langle \sqrt{s} \rangle$ = 197 GeV for a) b-events, b) non-b events, c) all events and d) the ratio between b- and non-b events compared to several QCD models.** The error bars include both statistical and systematic uncertainties.

• the W-boson pair-production background is estimated from the KORALW Monte Carlo and subtracted from the data, while releasing the cut on 4-jet events which are no longer removed from the data;

• the contamination from wrong-flavour events is estimated by varying the cut on the $B_n$ discriminant used to tag b events from 3.4 to 3.0 or 3.8 and the cut used to tag non-b events from 1.0 to 0.9 or 1.1. An additional lower cut at 0.2 is also introduced.

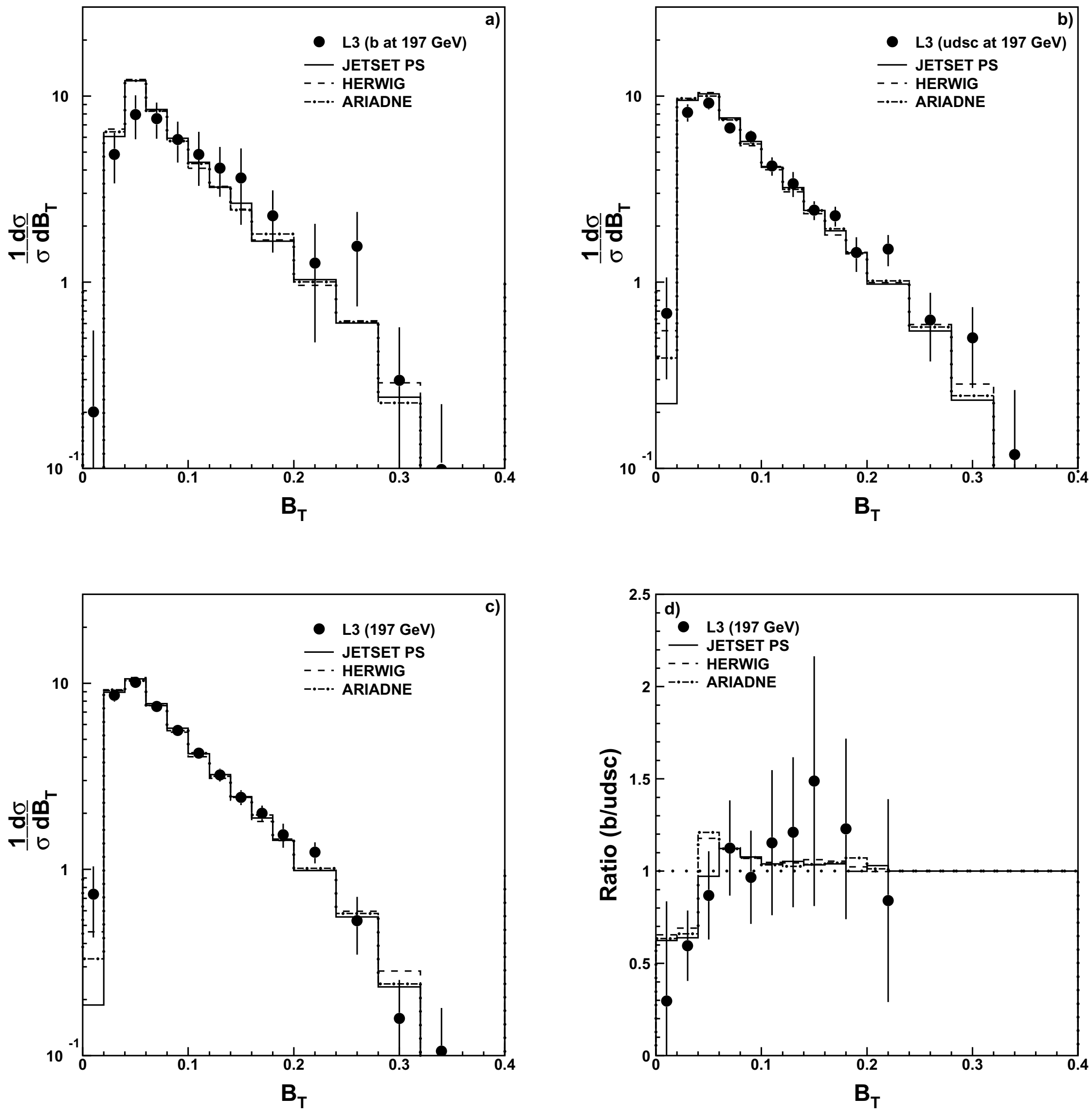

**Figure 6**

**Total jet broadening distributions at $\langle \sqrt{s} \rangle$ = 197 GeV for a) b-events, b) non-b events, c) all events and d) the ratio between b- and non-b events compared to several QCD models.** The error bars include both statistical and systematic uncertainties.

The bin-averaged systematic uncertainties due to different sources are summarized in Table 2 for the six event shape variables. Uncertainties due to detector corrections are between 4.8% and 6.0%, roughly 2–3 times larger than the uncertainty due to background estimation. The latter are dominated in equal parts by uncertainties due to radiative return and W-boson pair-production. In the flavour-tagged cases, the background uncertainty contains a significant contribution due to contamination from the wrong flavour and sometimes become the dominant source of systematic uncertainty. This uncertainty is between 2%–3% for the non-b events and 3%–10% for b-events.

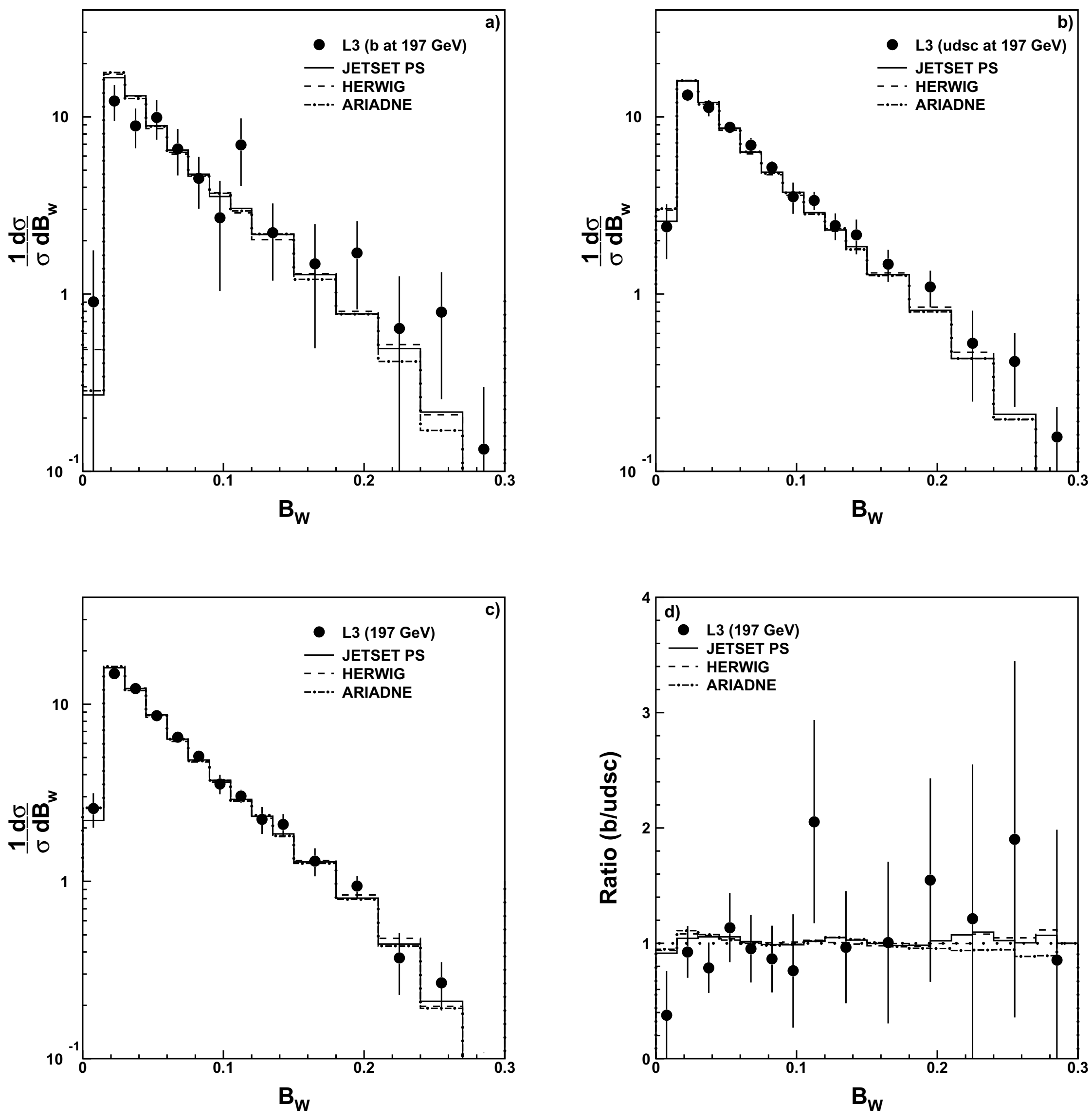

**Figure 7**

**Wide jet broadening distributions at $\langle\sqrt{s}\rangle$ = 197 GeV for a) b-events, b) non-b events, c) all events and d) the ratio between b- and non-b events compared to several QCD models.** The error bars include both statistical and systematic uncertainties.

The statistical component of the systematic uncertainty is negligible as the size of the Monte Carlo samples is at least 4 times, and sometimes even 10 times, larger than the size of the data sample. The final systematic uncertainty is taken as the sum in quadrature of all the contributions. Table 2 shows for each distribution the bin averaged systematic uncertainty as well as their contributions from different sources.

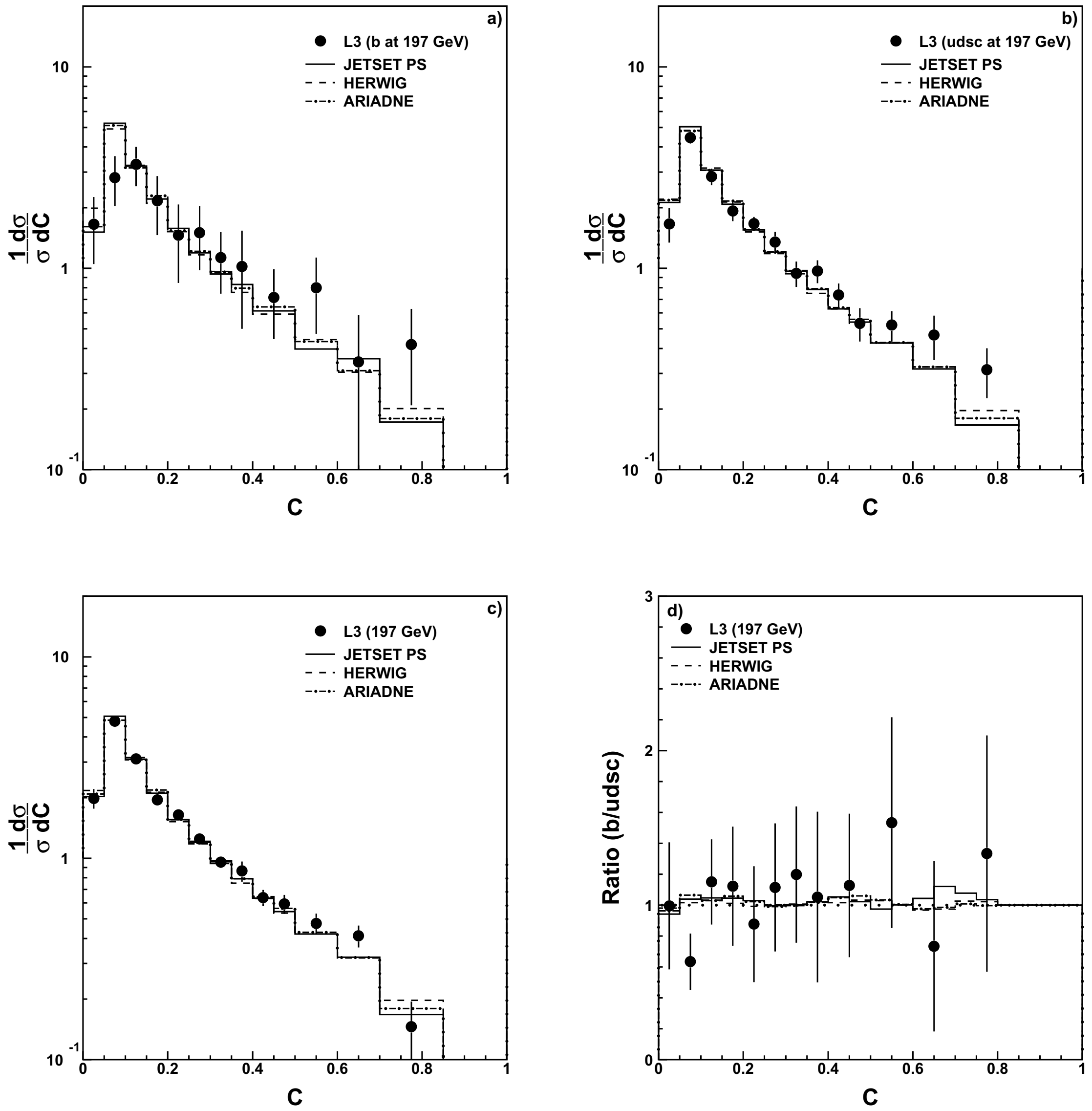

**Figure 8**

***C*-parameter distributions at $\langle\sqrt{s}\rangle$ = 197 GeV for a) b-events, b) non-b events, c) all events and d) the ratio between b- and non-b events compared to several QCD models.** The error bars include both statistical and systematic uncertainties.

## 8 Results

The corrected distributions for the six chosen event shape distributions, thrust, scaled heavy jet mass, total and wide jet broadening, *C*-parameter and 3-jet resolution parameter for the JADE algorithm, are summarized in Tables 3, 4, 5, 6, 7, 8 for $\langle\sqrt{s}\rangle$ = 197 GeV. These tables also show

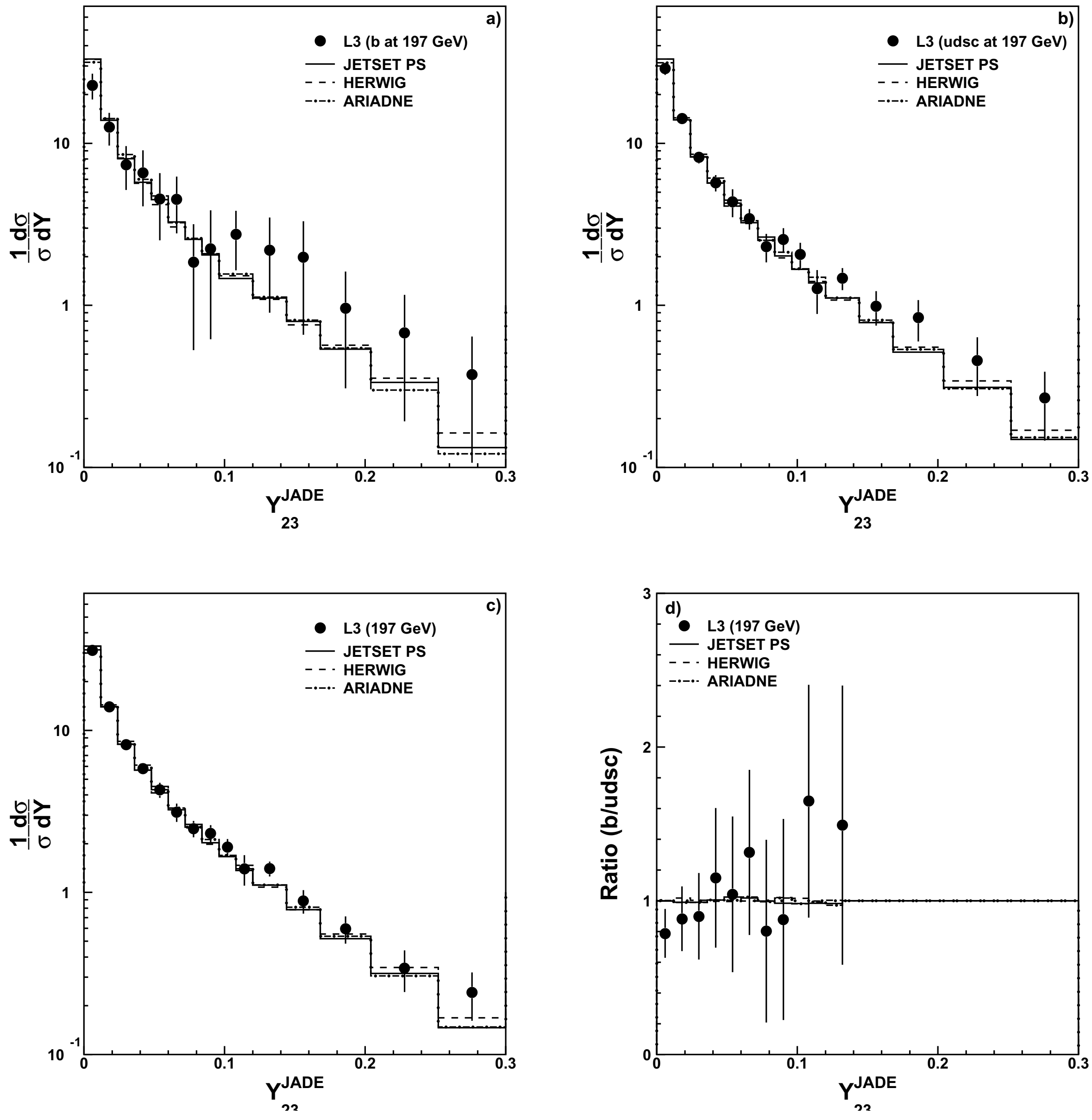

**Figure 9**

**Jet resolution parameter ($y^{J}_{23}$) distributions for 2 → 3 jet in JADE algorithm at $\langle \sqrt{s} \rangle$ = 197 GeV for a) b-events, b) non-b events, c) all events and d) the ratio between b- and non-b events compared to several QCD models.** The error bars include both statistical and systematic uncertainties.

the first and second moments of these distributions. The same six event shape distributions at $\sqrt{s}$ = 91.2 GeV were previously measured as reported in Reference [11].

**Table 2: Bin-averaged systematic uncertainties due to different sources for the six event shape variables at $\langle\sqrt{s}\rangle$ = 197 GeV for all, non-b and b events.**

| Event Sample | Source | $T$ | $\rho_H$ | $B_T$ | $B_W$ | $C$ | $y_{23}$ |
|---|---|---|---|---|---|---|---|
| All events | Detector | 5.6% | 5.9% | 4.8% | 6.6% | 5.5% | 6.0% |
| | Frag. Model | 0.6% | 1.3% | 1.5% | 1.4% | 1.6% | 0.5% |
| | Background | 2.2% | 2.3% | 2.6% | 2.4% | 2.4% | 2.3% |
| | Total | 6.2% | 6.8% | 6.1% | 7.6% | 6.4% | 6.7% |
| Non-b events | Detector | 5.9% | 7.4% | 5.5% | 7.3% | 6.9% | 7.4% |
| | Frag. Model | 0.9% | 1.4% | 1.1% | 1.1% | 1.3% | 0.4% |
| | Background | 2.6% | 2.7% | 3.7% | 3.0% | 3.2% | 3.1% |
| | Wrong Flavour | 1.8% | 2.1% | 2.0% | 3.0% | 2.3% | 2.8% |
| | Total | 7.1% | 8.6% | 7.2% | 9.0% | 8.9% | 8.5% |
| b events | Detector | 5.3% | 8.1% | 5.7% | 7.1% | 10.2% | 5.7% |
| | Frag. Model | 0.3% | 0.6% | 1.5% | 1.5% | 1.2% | 0.3% |
| | Background | 5.9% | 5.6% | 4.5% | 5.3% | 5.2% | 5.0% |
| | Wrong Flavour | 2.3% | 3.0% | 8.9% | 9.6% | 7.6% | 5.8% |
| | Total | 8.3% | 10.1% | 11.3% | 12.4% | 14.2% | 8.2% |

Figures 4, 5, 6, 7, 8, 9 show comparisons between data at $\langle\sqrt{s}\rangle$ = 197 GeV and predictions of the JETSET, ARIADNE and HERWIG models for distributions of thrust, scaled heavy jet mass, total

**Table 3: Differential distribution and first and second moments for event thrust at $\langle\sqrt{s}\rangle$ = 197 GeV for all, non-b and b events.**

| Thrust ($T$) | $\frac{1}{\sigma}\cdot\frac{d\sigma}{dT}$ (All) | $\frac{1}{\sigma}\cdot\frac{d\sigma}{dT}$ (Non-b) | Thrust ($T$) | $\frac{1}{\sigma}\cdot\frac{d\sigma}{dT}$ (b) |
|---|---|---|---|---|
| 0.500–0.600 | 0.00 ± 0.00 ± 0.00 | 0.00 ± 0.00 ± 0.00 | 0.500–0.600 | 0.00 ± 0.00 ± 0.00 |
| 0.600–0.650 | 0.01 ± 0.01 ± 0.02 | 0.01 ± 0.01 ± 0.04 | 0.600–0.650 | 0.00 ± 0.00 ± 0.00 |
| 0.650–0.700 | 0.13 ± 0.06 ± 0.06 | 0.26 ± 0.12 ± 0.11 | 0.650–0.700 | 0.04 ± 0.04 ± 0.03 |
| 0.700–0.750 | 0.19 ± 0.06 ± 0.04 | 0.42 ± 0.13 ± 0.15 | 0.700–0.750 | 0.63 ± 0.31 ± 0.36 |
| 0.750–0.800 | 0.56 ± 0.08 ± 0.11 | 0.67 ± 0.15 ± 0.15 | 0.750–0.800 | 0.62 ± 0.43 ± 0.37 |
| 0.800–0.825 | 0.80 ± 0.10 ± 0.11 | 0.88 ± 0.18 ± 0.18 | 0.800–0.850 | 1.14 ± 0.45 ± 0.59 |
| 0.825–0.850 | 1.05 ± 0.10 ± 0.08 | 1.23 ± 0.20 ± 0.18 | 0.850–0.900 | 2.95 ± 0.70 ± 0.44 |
| 0.850–0.875 | 1.62 ± 0.11 ± 0.17 | 1.76 ± 0.19 ± 0.26 | 0.900–0.925 | 3.43 ± 0.91 ± 1.09 |
| 0.875–0.900 | 1.72 ± 0.10 ± 0.21 | 1.60 ± 0.17 ± 0.32 | 0.925–0.950 | 5.02 ± 1.05 ± 0.45 |
| 0.900–0.925 | 3.03 ± 0.12 ± 0.23 | 3.24 ± 0.22 ± 0.19 | 0.950–0.975 | 8.97 ± 1.43 ± 0.97 |
| 0.925–0.950 | 4.72 ± 0.14 ± 0.23 | 5.09 ± 0.27 ± 0.38 | 0.975–1.000 | 11.83 ± 1.94 ± 0.70 |
| 0.950–0.975 | 9.24 ± 0.19 ± 0.22 | 8.95 ± 0.37 ± 0.24 | | |
| 0.975–1.000 | 16.04 ± 0.25 ± 1.09 | 14.54 ± 0.56 ± 1.11 | | |
| First Moment | 0.943 ± 0.010 ± 0.004 | 0.935 ± 0.020 ± 0.003 | | 0.927 ± 0.072 ± 0.010 |
| Second Moment | 0.893 ± 0.010 ± 0.007 | 0.879 ± 0.021 ± 0.006 | | 0.865 ± 0.072 ± 0.016 |

The first and the second errors refer to statistical and systematic uncertainties respectively.

**Table 4: Differential distribution and first and second moments for scaled heavy jet mass at $\left\langle \sqrt{s} \right\rangle$ = 197 GeV for all, non-b and b events.**

| $\rho_H$ | $\frac{1}{\sigma} \cdot \frac{d\sigma}{d\rho_H}$ (All) | $\frac{1}{\sigma} \cdot \frac{d\sigma}{d\rho_H}$ (Non-b) | $\rho_H$ | $\frac{1}{\sigma} \cdot \frac{d\sigma}{d\rho_H}$ (b) |
|---|---|---|---|---|
| 0.000–0.015 | 20.31 ± 0.33 ± 1.68 | 18.24 ± 0.74 ± 1.76 | 0.000–0.015 | 15.10 ± 2.49 ± 1.16 |
| 0.015–0.030 | 15.93 ± 0.34 ± 0.61 | 15.60 ± 0.69 ± 0.73 | 0.015–0.030 | 15.32 ± 2.71 ± 0.95 |
| 0.030–0.045 | 8.72 ± 0.26 ± 0.19 | 8.58 ± 0.47 ± 0.44 | 0.030–0.045 | 7.77 ± 1.93 ± 1.72 |
| 0.045–0.060 | 5.12 ± 0.21 ± 0.41 | 5.18 ± 0.38 ± 0.56 | 0.045–0.060 | 4.95 ± 1.46 ± 1.11 |
| 0.060–0.075 | 3.59 ± 0.18 ± 0.42 | 3.74 ± 0.33 ± 0.66 | 0.060–0.075 | 4.39 ± 1.39 ± 0.45 |
| 0.075–0.090 | 2.76 ± 0.17 ± 0.14 | 3.05 ± 0.31 ± 0.14 | 0.075–0.090 | 4.34 ± 1.57 ± 0.51 |
| 0.090–0.105 | 2.22 ± 0.16 ± 0.27 | 2.37 ± 0.28 ± 0.42 | 0.090–0.120 | 3.46 ± 0.91 ± 0.53 |
| 0.105–0.120 | 1.89 ± 0.16 ± 0.25 | 1.96 ± 0.28 ± 0.32 | 0.120–0.150 | 1.69 ± 0.60 ± 0.97 |
| 0.120–0.150 | 1.11 ± 0.10 ± 0.12 | 1.22 ± 0.18 ± 0.16 | 0.150–0.180 | 0.95 ± 0.69 ± 0.45 |
| 0.150–0.180 | 0.75 ± 0.10 ± 0.06 | 0.96 ± 0.18 ± 0.16 | 0.180–0.210 | 0.49 ± 0.40 ± 0.49 |
| 0.180–0.210 | 0.51 ± 0.09 ± 0.11 | 0.63 ± 0.17 ± 0.17 | 0.210–0.240 | 0.21 ± 0.26 ± 0.21 |
| 0.210–0.240 | 0.27 ± 0.08 ± 0.06 | 0.39 ± 0.14 ± 0.13 | 0.240–0.270 | 0.32 ± 0.23 ± 0.17 |
| 0.240–0.270 | 0.23 ± 0.07 ± 0.04 | 0.41 ± 0.15 ± 0.11 | 0.270–0.300 | 0.26 ± 0.15 ± 0.10 |
| 0.270–0.300 | 0.17 ± 0.06 ± 0.05 | 0.37 ± 0.16 ± 0.14 | | |
| First Moment | 0.046 ± 0.001 ± 0.003 | 0.053 ± 0.002 ± 0.003 | | 0.057 ± 0.005 ± 0.006 |
| Second Moment | 0.005 ± 0.001 ± 0.001 | 0.006 ± 0.001 ± 0.001 | | 0.006 ± 0.001 ± 0.001 |

The first and the second errors refer to statistical and systematic uncertainties respectively.

and wide jet broadening, *C*-parameter and the 3-jet JADE resolution parameter for all hadronic events, b-events and non-b events. The error bars shown in these figures are the quadratic sum of

**Table 5: Differential distribution and first and second moments for total jet broadening at $\left\langle \sqrt{s} \right\rangle$ = 197 GeV for all, non-b and b events.**

| $B_T$ | $\frac{1}{\sigma} \cdot \frac{d\sigma}{dB_T}$ (All) | $\frac{1}{\sigma} \cdot \frac{d\sigma}{dB_T}$ (Non-b) | $B_T$ | $\frac{1}{\sigma} \cdot \frac{d\sigma}{dB_T}$ (b) |
|---|---|---|---|---|
| 0.000–0.020 | 0.75 ± 0.06 ± 0.30 | 0.68 ± 0.11 ± 0.37 | 0.000–0.020 | 0.20 ± 0.20 ± 0.29 |
| 0.020–0.040 | 8.61 ± 0.21 ± 0.61 | 8.17 ± 0.50 ± 0.71 | 0.020–0.040 | 4.80 ± 1.30 ± 0.68 |
| 0.040–0.060 | 10.10 ± 0.22 ± 0.41 | 9.15 ± 0.48 ± 0.51 | 0.040–0.060 | 8.03 ± 1.84 ± 1.08 |
| 0.060–0.080 | 7.50 ± 0.19 ± 0.16 | 6.73 ± 0.37 ± 0.18 | 0.060–0.080 | 7.47 ± 1.53 ± 0.64 |
| 0.080–0.100 | 5.58 ± 0.16 ± 0.18 | 6.03 ± 0.33 ± 0.37 | 0.080–0.100 | 5.85 ± 1.27 ± 0.70 |
| 0.100–0.120 | 4.17 ± 0.15 ± 0.18 | 4.18 ± 0.27 ± 0.39 | 0.100–0.120 | 4.83 ± 1.26 ± 0.91 |
| 0.120–0.140 | 3.22 ± 0.13 ± 0.22 | 3.37 ± 0.24 ± 0.46 | 0.120–0.140 | 4.15 ± 1.12 ± 0.54 |
| 0.140–0.160 | 2.43 ± 0.12 ± 0.19 | 2.43 ± 0.23 ± 0.17 | 0.140–0.160 | 3.62 ± 1.05 ± 1.20 |
| 0.160–0.180 | 2.01 ± 0.12 ± 0.16 | 2.27 ± 0.22 ± 0.16 | 0.160–0.200 | 2.27 ± 0.67 ± 0.47 |
| 0.180–0.200 | 1.54 ± 0.12 ± 0.19 | 1.46 ± 0.22 ± 0.22 | 0.200–0.240 | 1.28 ± 0.63 ± 0.49 |
| 0.200–0.240 | 1.24 ± 0.10 ± 0.12 | 1.52 ± 0.19 ± 0.22 | 0.240–0.280 | 1.58 ± 0.72 ± 0.40 |
| 0.240–0.280 | 0.53 ± 0.10 ± 0.15 | 0.63 ± 0.18 ± 0.17 | 0.280–0.320 | 0.30 ± 0.21 ± 0.21 |
| 0.280–0.320 | 0.16 ± 0.07 ± 0.07 | 0.50 ± 0.14 ± 0.18 | 0.320–0.360 | 0.10 ± 0.10 ± 0.07 |
| 0.320–0.360 | 0.11 ± 0.05 ± 0.06 | 0.12 ± 0.07 ± 0.13 | | |
| 0.360–0.400 | 0.01 ± 0.01 ± 0.01 | 0.01 ± 0.01 ± 0.01 | | |
| First Moment | 0.093 ± 0.001 ± 0.004 | 0.100 ± 0.002 ± 0.004 | | 0.114 ± 0.007 ± 0.008 |
| Second Moment | 0.013 ± 0.001 ± 0.001 | 0.015 ± 0.001 ± 0.001 | | 0.018 ± 0.002 ± 0.003 |

The first and the second errors refer to statistical and systematic uncertainties respectively.

**Table 6: Differential distribution and first and second moments for wide jet broadening at $\langle\sqrt{s}\rangle$ = 197 GeV for all, non-b and b events.**

| $B_W$ | $\frac{1}{\sigma}\cdot\frac{d\sigma}{dB_W}$ (All) | $\frac{1}{\sigma}\cdot\frac{d\sigma}{dB_W}$ (Non-b) | $B_W$ | $\frac{1}{\sigma}\cdot\frac{d\sigma}{dB_W}$ (b) |
|---|---|---|---|---|
| 0.000–0.015 | 2.57 ± 0.12 ± 0.55 | 2.39 ± 0.27 ± 0.78 | 0.000–0.015 | 1.02 ± 0.51 ± 0.73 |
| 0.015–0.030 | 14.86 ± 0.32 ± 0.92 | 13.29 ± 0.72 ± 0.76 | 0.015–0.030 | 12.33 ± 2.56 ± 1.19 |
| 0.030–0.045 | 12.25 ± 0.27 ± 0.80 | 11.28 ± 0.59 ± 1.07 | 0.030–0.045 | 8.81 ± 1.91 ± 1.18 |
| 0.045–0.060 | 8.61 ± 0.24 ± 0.27 | 8.74 ± 0.49 ± 0.43 | 0.045–0.060 | 9.78 ± 2.00 ± 1.45 |
| 0.060–0.075 | 6.49 ± 0.21 ± 0.25 | 6.87 ± 0.42 ± 0.49 | 0.060–0.075 | 6.70 ± 1.72 ± 0.95 |
| 0.075–0.090 | 5.06 ± 0.20 ± 0.27 | 5.19 ± 0.36 ± 0.19 | 0.075–0.090 | 4.46 ± 1.30 ± 0.63 |
| 0.090–0.105 | 3.53 ± 0.17 ± 0.40 | 3.53 ± 0.31 ± 0.63 | 0.090–0.105 | 2.68 ± 0.98 ± 1.33 |
| 0.105–0.120 | 3.03 ± 0.17 ± 0.19 | 3.36 ± 0.31 ± 0.25 | 0.105–0.120 | 7.14 ± 1.87 ± 2.32 |
| 0.120–0.135 | 2.24 ± 0.16 ± 0.36 | 2.45 ± 0.29 ± 0.31 | 0.120–0.150 | 2.19 ± 0.74 ± 0.67 |
| 0.135–0.150 | 2.10 ± 0.16 ± 0.26 | 2.16 ± 0.30 ± 0.38 | 0.150–0.180 | 1.50 ± 0.64 ± 0.77 |
| 0.150–0.180 | 1.31 ± 0.11 ± 0.21 | 1.49 ± 0.19 ± 0.24 | 0.180–0.210 | 1.68 ± 0.67 ± 0.55 |
| 0.180–0.210 | 0.94 ± 0.11 ± 0.07 | 1.10 ± 0.19 ± 0.17 | 0.210–0.240 | 0.59 ± 0.49 ± 0.25 |
| 0.210–0.240 | 0.37 ± 0.10 ± 0.10 | 0.53 ± 0.17 ± 0.22 | 0.240–0.270 | 0.79 ± 0.39 ± 0.36 |
| 0.240–0.270 | 0.27 ± 0.07 ± 0.04 | 0.42 ± 0.14 ± 0.13 | 0.270–0.300 | 0.13 ± 0.13 ± 0.10 |
| 0.270–0.300 | 0.07 ± 0.03 ± 0.03 | 0.16 ± 0.06 ± 0.04 | | |
| First Moment | 0.068 ± 0.001 ± 0.003 | 0.073 ± 0.002 ± 0.003 | | 0.083 ± 0.005 ± 0.006 |
| Second Moment | 0.007 ± 0.001 ± 0.001 | 0.009 ± 0.001 ± 0.001 | | 0.010 ± 0.001 ± 0.001 |

The first and the second errors refer to statistical and systematic uncertainties respectively.

statistical and systematic uncertainties. The ratios of the event shape distributions for b- and non-b events are also shown together with predictions from parton shower models. For the b-events

**Table 7: Differential distribution and first and second moments for C-parameter at $\langle\sqrt{s}\rangle$ = 197 GeV for all, non-b and b events.**

| C-parameter | $\frac{1}{\sigma}\cdot\frac{d\sigma}{dC}$ (All) | $\frac{1}{\sigma}\cdot\frac{d\sigma}{dC}$ (Non-b) | C-parameter | $\frac{1}{\sigma}\cdot\frac{d\sigma}{dC}$ (b) |
|---|---|---|---|---|
| 0.000–0.050 | 1.98 ± 0.07 ± 0.21 | 1.66 ± 0.13 ± 0.29 | 0.000–0.050 | 1.68 ± 0.48 ± 0.37 |
| 0.050–0.100 | 4.80 ± 0.10 ± 0.29 | 4.45 ± 0.22 ± 0.23 | 0.050–0.100 | 2.82 ± 0.68 ± 0.40 |
| 0.100–0.150 | 3.10 ± 0.08 ± 0.13 | 2.84 ± 0.16 ± 0.22 | 0.100–0.150 | 3.21 ± 0.64 ± 0.29 |
| 0.150–0.200 | 1.95 ± 0.06 ± 0.07 | 1.92 ± 0.12 ± 0.18 | 0.150–0.200 | 2.18 ± 0.52 ± 0.48 |
| 0.200–0.250 | 1.64 ± 0.06 ± 0.06 | 1.66 ± 0.11 ± 0.08 | 0.200–0.250 | 1.45 ± 0.42 ± 0.43 |
| 0.250–0.300 | 1.23 ± 0.05 ± 0.05 | 1.34 ± 0.10 ± 0.14 | 0.250–0.300 | 1.49 ± 0.41 ± 0.32 |
| 0.300–0.350 | 0.97 ± 0.05 ± 0.04 | 0.96 ± 0.08 ± 0.11 | 0.300–0.350 | 1.12 ± 0.33 ± 0.19 |
| 0.350–0.400 | 0.85 ± 0.05 ± 0.09 | 0.95 ± 0.09 ± 0.09 | 0.350–0.400 | 1.04 ± 0.36 ± 0.40 |
| 0.400–0.450 | 0.64 ± 0.04 ± 0.04 | 0.74 ± 0.08 ± 0.07 | 0.400–0.500 | 0.72 ± 0.23 ± 0.16 |
| 0.450–0.500 | 0.59 ± 0.04 ± 0.05 | 0.53 ± 0.07 ± 0.07 | 0.500–0.600 | 0.80 ± 0.26 ± 0.18 |
| 0.500–0.600 | 0.48 ± 0.03 ± 0.05 | 0.53 ± 0.06 ± 0.07 | 0.600–0.700 | 0.34 ± 0.19 ± 0.14 |
| 0.600–0.700 | 0.41 ± 0.04 ± 0.03 | 0.47 ± 0.07 ± 0.09 | 0.700–0.850 | 0.44 ± 0.16 ± 0.14 |
| 0.700–0.850 | 0.15 ± 0.03 ± 0.03 | 0.31 ± 0.07 ± 0.05 | | |
| 0.850–1.000 | 0.01 ± 0.01 ± 0.01 | 0.01 ± 0.01 ± 0.01 | | |
| First Moment | 0.222 ± 0.004 ± 0.014 | 0.248 ± 0.007 ± 0.011 | | 0.271 ± 0.019 ± 0.028 |
| Second Moment | 0.084 ± 0.003 ± 0.010 | 0.104 ± 0.006 ± 0.008 | | 0.117 ± 0.015 ± 0.026 |

The first and the second errors refer to statistical and systematic uncertainties respectively.

**Table 8: Differential distribution and first and second moments for 3-jet resolution parameter ($y_{23}^{J}$) in Jade algorithm at $\langle \sqrt{s} \rangle$ = 197 GeV for all, non-b and b events.**

| $y_{23}^{J}$ | $\frac{1}{\sigma} \cdot \frac{d\sigma}{dy_{23}^{J}}$ (All) | $\frac{1}{\sigma} \cdot \frac{d\sigma}{dy_{23}^{J}}$ (Non-b) | $y_{23}^{J}$ | $\frac{1}{\sigma} \cdot \frac{d\sigma}{dy_{23}^{J}}$ (b) |
|---|---|---|---|---|
| 0.000–0.012 | 37.89 ± 0.62 ± 3.70 | 33.51 ± 1.30 ± 3.25 | 0.000–0.012 | 29.89 ± 4.75 ± 4.44 |
| 0.012–0.024 | 12.82 ± 0.31 ± 0.75 | 13.47 ± 0.65 ± 0.81 | 0.012–0.024 | 11.72 ± 2.48 ± 0.96 |
| 0.024–0.036 | 7.04 ± 0.24 ± 0.56 | 7.34 ± 0.47 ± 0.58 | 0.024–0.036 | 6.41 ± 1.68 ± 1.03 |
| 0.036–0.048 | 4.93 ± 0.20 ± 0.47 | 5.18 ± 0.40 ± 0.54 | 0.036–0.048 | 6.03 ± 1.62 ± 1.66 |
| 0.048–0.060 | 3.63 ± 0.18 ± 0.49 | 3.82 ± 0.35 ± 0.74 | 0.048–0.060 | 4.07 ± 1.24 ± 1.15 |
| 0.060–0.072 | 2.73 ± 0.16 ± 0.35 | 3.21 ± 0.32 ± 0.37 | 0.060–0.072 | 4.19 ± 1.33 ± 0.95 |
| 0.072–0.084 | 2.12 ± 0.15 ± 0.26 | 2.20 ± 0.28 ± 0.32 | 0.072–0.084 | 1.54 ± 0.71 ± 0.55 |
| 0.084–0.096 | 2.01 ± 0.15 ± 0.24 | 2.29 ± 0.27 ± 0.34 | 0.084–0.096 | 1.91 ± 1.04 ± 0.57 |
| 0.096–0.108 | 1.66 ± 0.14 ± 0.20 | 1.97 ± 0.27 ± 0.24 | 0.096–0.120 | 2.22 ± 0.74 ± 0.62 |
| 0.108–0.120 | 1.14 ± 0.13 ± 0.31 | 1.14 ± 0.23 ± 0.30 | 0.120–0.144 | 1.67 ± 0.68 ± 0.80 |
| 0.120–0.144 | 1.26 ± 0.10 ± 0.11 | 1.39 ± 0.18 ± 0.13 | 0.144–0.168 | 1.74 ± 0.74 ± 0.75 |
| 0.144–0.168 | 0.69 ± 0.09 ± 0.11 | 0.82 ± 0.17 ± 0.15 | 0.168–0.204 | 0.89 ± 0.45 ± 0.50 |
| 0.168–0.204 | 0.47 ± 0.08 ± 0.08 | 0.72 ± 0.14 ± 0.18 | 0.204–0.252 | 0.55 ± 0.31 ± 0.22 |
| 0.204–0.252 | 0.31 ± 0.06 ± 0.06 | 0.41 ± 0.12 ± 0.10 | 0.252–0.300 | 0.35 ± 0.19 ± 0.15 |
| 0.252–0.300 | 0.21 ± 0.05 ± 0.04 | 0.24 ± 0.08 ± 0.08 | | |
| First Moment | 0.044 ± 0.001 ± 0.003 | 0.048 ± 0.002 ± 0.003 | | 0.060 ± 0.006 ± 0.008 |
| Second Moment | 0.005 ± 0.001 ± 0.001 | 0.006 ± 0.001 ± 0.001 | | 0.008 ± 0.001 ± 0.002 |

The first and the second errors refer to statistical and systematic uncertainties respectively.

in the two-jet region, the model predictions seem to overestimate the data, in particular for the thrust (Figure 4a), wide jet broadening (Figure 7a) and *C*-parameter (Figure 8a) distributions.

The agreement between the three models with the data is quantified in Table 9 which summarizes the $\chi^2$ and the confidence level of a comparison of these models with the data for the six event-shape variables for the three data samples. An overall good agreement between data and the model predictions is observed. All three models describe equally well the data, the minimum confidence level being 0.11 for the HERWIG comparison with $B_W$ for non-b events. The overall agreement obtained for the three distributions singled out above presenting local discrepancies for b-events in the two-jet region is found to be quite satisfactory.

Since the models were tuned only on low energy data and on all, or only udsc, quark flavours, the agreement observed shows that the energy evolution of QCD processes in the range between 90 GeV and 200 GeV, as well as the production of b quarks, is correctly described by the models considered. The event shape variables considered are, however, not very sensitive to differences between heavy and light quarks. Only in the distributions of $B_T$, at low values (Figure 6d) does the ratio of b to non-b events depart markedly from unity, a feature that is correctly described by the models.

**Table 9: Comparison of different parton shower models with the data at $\langle \sqrt{s} \rangle$ = 197 GeV for all events, non-b events and b events for the six event-shape variables.**

| Event Sample | Model | | $T$ | $\rho_H$ | $B_T$ | $B_W$ | C | $y_{23}$ |
|---|---|---|---|---|---|---|---|---|
| All events | Jetset | $\chi^2$/d.o.f. | 7.7/12 | 6.9/14 | 10.2/15 | 7.4/15 | 10.4/14 | 9.7/15 |
| | | C.L. | 0.74 | 0.91 | 0.75 | 0.92 | 0.66 | 0.79 |
| | Herwig | $\chi^2$/d.o.f. | 9.0/12 | 8.5/14 | 10.1/15 | 9.9/15 | 14.9/14 | 9.7/15 |
| | | C.L. | 0.62 | 0.81 | 0.75 | 0.77 | 0.32 | 0.78 |
| | Ariadne | $\chi^2$/d.o.f. | 6.9/12 | 7.6/14 | 6.4/15 | 9.0/15 | 12.5/14 | 9.7/15 |
| | | C.L. | 0.80 | 0.87 | 0.95 | 0.83 | 0.48 | 0.78 |
| Non-b events | Jetset | $\chi^2$/d.o.f. | 15.1/12 | 12.3/14 | 20.1/15 | 17.5/15 | 16.8/14 | 12.6/15 |
| | | C.L. | 0.18 | 0.50 | 0.13 | 0.23 | 0.21 | 0.56 |
| | Herwig | $\chi^2$/d.o.f. | 15.1/12 | 12.3/14 | 20.1/15 | 20.5/15 | 17.0/14 | 12.3/15 |
| | | C.L. | 0.18 | 0.50 | 0.13 | 0.11 | 0.20 | 0.58 |
| | Ariadne | $\chi^2$/d.o.f. | 13.4/12 | 12.9/14 | 16.1/15 | 19.7/15 | 14.8/14 | 10.3/15 |
| | | C.L. | 0.27 | 0.46 | 0.31 | 0.14 | 0.32 | 0.74 |
| b events | Jetset | $\chi^2$/d.o.f. | 11.1/9 | 11.6/13 | 12.8/13 | 11.8/14 | 12.5/12 | 12.1/14 |
| | | C.L. | 0.20 | 0.48 | 0.38 | 0.55 | 0.33 | 0.52 |
| | Herwig | $\chi^2$/d.o.f. | 11.5/9 | 10.6/13 | 14.5/13 | 13.7/14 | 11.0/12 | 11.9/14 |
| | | C.L. | 0.18 | 0.56 | 0.27 | 0.40 | 0.45 | 0.54 |
| | Ariadne | $\chi^2$/d.o.f. | 10.0/9 | 11.0/13 | 13.6/13 | 13.4/14 | 10.7/12 | 10.1/14 |
| | | C.L. | 0.27 | 0.53 | 0.32 | 0.42 | 0.47 | 0.68 |

The $\chi^2$ over the numbers of degrees of freedom (d.o.f.) and the confidence levels are shown.

## 9 Summary

Event shape distributions for hadronic events are studied from $e^+e^-$ annihilation data collected by the L3 detector at LEP at $\langle \sqrt{s} \rangle$ = 197 GeV. Flavour tagging is used to separate a b-quark enriched sample from a sample of lighter flavours.

The event shape distributions are well described by all the parton shower models JETSET, HERWIG and ARIADNE.

## L3 Collaboration

P.Achard[20], O. Adriani[17], M. Aguilar-Benitez[25], J. Alcaraz[25], G. Alemanni[23], J. Allaby[18], A. Aloisio[29], M. G. Alviggi[29], H. Anderhub[49], V. P. Andreev[6,34], F. Anselmo[8], A. Arefiev[28], T. Azemoon[3], T. Aziz[9], P. Bagnaia[39], A. Bajo[25], G. Baksay[26], L. Baksay[26], S. V. Baldew[2], S. Banerjee[9], Sw. Banerjee[4], A. Barczyk[49,47], R. Barillère[18], P. Bartalini[23], M. Basile[8], N. Batalova[46], R. Battiston[33], A. Bay[23], U. Becker[13], F. Behner[49], L. Bellucci[17], R. Berbeco[3], J. Berdugo[25], P. Berges[13], B. Bertucci[33], B. L. Betev[49], M. Biasini[33], M. Biglietti[29], A. Biland[49], J. J. Blaising[4], S. C. Blyth[35], G. J. Bobbink[2], A. Böhm[1], L. Boldizsar[12], B. Borgia[39], S. Bottai[17], D. Bourilkov[49], M. Bourquin[20], S. Braccini[20], J. G. Branson[41], F. Brochu[4], J. D. Burger[13], W. J. Burger[33], X. D. Cai[13], M. Capell[13], G. Cara Romeo[8], G. Carlino[29], A. Cartacci[17], J. Casaus[25], F. Cavallari[39], N. Cavallo[36], C. Cecchi[33], M. Cerrada[25], M. Chamizo[20], Y. H. Chang[44], M. Chemarin[24], A. Chen[44], G. Chen[7], G. M. Chen[7], H. F. Chen[22], H. S. Chen[7], G. Chiefari[29], L. Cifarelli[40], F. Cindolo[8], I. Clare[13], R. Clare[38], G. Coignet[4], N. Colino[25], S. Costantini[39], B. de la Cruz[25], S. Cucciarelli[33], R. de Asmundis[29], P. Déglon[20], J. Debreczeni[12], A. Degré[4], K. Dehmelt[26], K. Deiters[47], D. della Volpe[29], E. Delmeire[20], P. Denes[37], F. DeNotaristefani[39], A. De Salvo[49], M. Diemoz[39], M. Dierckxsens[2], C. Dionisi[39], M. Dittmar[49], A. Doria[29], M. T. Dova[10], D. Duchesneau[4], M. Duda[1], B. Echenard[20], A. Eline[18], A. El Hage[1], H. El Mamouni[24], A. Engler[35], F. J. Eppling[13], P. Extermann[20], M. A. Falagan[25], S. Falciano[39], A. Favara[32], J. Fay[24], O. Fedin[34], M. Felcini[49], T. Ferguson[35], H. Fesefeldt[1], E. Fiandrini[33], J. H. Field[20], F. Filthaut[31], P. H. Fisher[13], W. Fisher[37], G. Forconi[13], K. Freudenreich[49], C. Furetta[27], Yu. Galaktionov[28,13], S. N. Ganguli[9], P. Garcia-Abia[25], M. Gataullin[32], S. Gentile[39], S. Giagu[39], Z. F. Gong[22], G. Grenier[24], O. Grimm[49], M. W. Gruenewald[16], V. K. Gupta[37], A. Gurtu[9], L. J. Gutay[46], D. Haas[5], D. Hatzifotiadou[8], T. Hebbeker[1], A. Hervé[18], J. Hirschfelder[35], H. Hofer[49], M. Hohlmann[26], G. Holzner[49], S. R. Hou[44], B. N. Jin[7], P. Jindal[14], L. W. Jones[4], P. de Jong[2], I. Josa-Mutuberria[25], M. Kaur[14], M. N. Kienzle-Focacci[20], J. K. Kim[43], J. Kirkby[18], W. Kittel[31], A. Klimentov[13,28], A. C. König[31], M. Kopal[46], V. Koutsenko[13,28], M. Kräber[49], R. W. Kraemer[35], A.

Krüger[48], A. Kunin[13], P. Ladron de Guevara[25], I. Laktineh[24], G. Landi[17], M. Lebeau[18], A. Lebedev[13], P. Lebrun[24], P. Lecomte[49], P. Lecoq[18], P. Le Coultre[49], J. M. Le Goff[18], R. Leiste[48], M. Levtchenko[27], P. Levtchenko[34], C. Li[22], S. Likhoded[48], C. H. Lin[44], W. T. Lin[44], F. L. Linde[2], L. Lista[29], Z. A. Liu[7], W. Lohmann[48], E. Longo[39], Y. S. Lu[7], C. Luci[39], L. Luminari[39], W. Lustermann[49], W. G. Ma[22], L. Malgeri[18], A. Malinin[28], C. Maña[25], J. Mans[37], J. P. Martin[24], F. Marzano[39], K. Mazumdar[9], R. R. McNeil[6], S. Mele[18,29] Salvatore.Mele@cern.ch, L. Merola[29], M. Meschini[17], W. J. Metzger[31], A. Mihul[11], H. Milcent[18], G. Mirabelli[39], J. Mnich[1], G. B. Mohanty[9], G. S. Muanza[24], A. J. M. Muijs[2], M. Musy[39], S. Nagy[15], S. Natale[20], M. Napolitano[29], F. Nessi-Tedaldi[49], H. Newman[32], A. Nisati[39], T. Novak[31], H. Nowak[48], R. Ofierzynski[49], G. Organtini[39], I. Pal[46], C. Palomares[25], P. Paolucci[29], R. Paramatti[39], G. Passaleva[17], S. Patricelli[29], T. Paul[10], M. Pauluzzi[33], C. Paus[13], F. Pauss[49], M. Pedace[39], S. Pensotti[27], D. Perret-Gallix[4], D. Piccolo[29], F. Pierella[8], M. Pieri[41], M. Pioppi[33], P. A. Piroué[37], E. Pistolesi[27], V. Plyaskin[28], M. Pohl[20], V. Pojidaev[17], J. Pothier[18], D. Prokofiev[34], G. Rahal-Callot[49], M. A. Rahaman[9], P. Raics[15], N. Raja[9], R. Ramelli[49], P. G. Rancoita[27], R. Ranieri[17], A. Raspereza[48], P. Razis[30], S. Rembeczki[26], D. Ren[49], M. Rescigno[39], S. Reucroft[10], S. Riemann[48], K. Riles[3], B. P. Roe[3], L. Romero[25], A. Rosca[48], C. Rosemann[1], C. Rosenbleck[1], S. Rosier-Lees[4], S. Roth[1], J. A. Rubio[18], G. Ruggiero[17], H. Rykaczewski[49], A. Sakharov[49], S. Saremi[6], S. Sarkar[39], J. Salicio[18], E. Sanchez[25], C. Schäfer[18], V. Schegelsky[34], H. Schopper[21], D. J. Schotanus[31], C. Sciacca[29], L. Servoli[33], S. Shevchenko[32], N. Shivarov[42], V. Shoutko[13], E. Shumilov[28], A. Shvorob[32], D. Son[43], C. Souga[24], P. Spillantini[17], M. Steuer[13], D. P. Stickland[37], B. Stoyanov[42], A. Straessner[20], K. Sudhakar[9], G. Sultanov[42], L. Z. Sun[22], S. Sushkov[1], H. Suter[49], J. D. Swain[10], Z. Szillasi[26], X. W. Tang[7], P. Tarjan[15], L. Tauscher[5], L. Taylor[10], B. Tellili[24], D. Teyssier[24], C. Timmermans[31], Samuel. C. C. Ting[13], S. M. Ting[13], S. C. Tonwar[9], J. Tóth[12], C. Tully[37], K. L. Tung[7], J. Ulbricht[49], E. Valente[39], R. T. Van de Walle[31], R. Vasquez[46], G. Vesztergombi[12], I. Vetlitsky[28], G. Viertel[49], M. Vivargent[4], S. Vlachos[5], I. Vodopianov[26], H. Vogel[35], H. Vogt[48], I. Vorobiev[35,28], A. A. Vorobyov[34], M. Wadhwa[5], Q. Wang[31], X. L. Wang[22], Z. M. Wang[22], M. Weber[18], S. Wynhoff[37],‡, L. Xia[32], Z. Z. Xu[22], J. Yamamoto[3], B. Z. Yang[22], C. G. Yang[7], H. J. Yang[3], M. Yang[7], S. C. Yeh[45], An. Zalite[34], Yu. Zalite[34], Z. P. Zhang[22], J. Zhao[22], G. Y. Zhu[7], R. Y. Zhu[32], H. L. Zhuang[7], A. Zichichi[8,18,19], B. Zimmermann[49], M. Zöller[1]

[1]III. Physikalisches Institut, RWTH, D-52056 Aachen, Germany.

[2]National Institute for High Energy Physics, NIKHEF, and University of Amsterdam, NL-1009 DB Amsterdam, The Netherlands.

[3]University of Michigan, Ann Arbor, MI 48109, USA.

[4]Laboratoire d'Annecy-le-Vieux de Physique des Particules, LAPP, IN2P3-CNRS, BP 110, F-74941 Annecy-le-Vieux CEDEX, France.

[5]Institute of Physics, University of Basel, CH-4056 Basel, Switzerland.

[6]Louisiana State University, Baton Rouge, LA 70803, USA.

[7]Institute of High Energy Physics, IHEP, 100039 Beijing, China△.

[8]University of Bologna and INFN-Sezione di Bologna, I-40126 Bologna, Italy.

[9]Tata Institute of Fundamental Research, Mumbai (Bombay) 400 005, India.

[10]Northeastern University, Boston, MA 02115, USA.

[11]Institute of Atomic Physics and University of Bucharest, R-76900 Bucharest, Romania.

[12]Central Research Institute for Physics of the Hungarian Academy of Sciences, H-1525 Buda pest 114, Hungary[‡].

[13]Massachusetts Institute of Technology, Cambridge, MA 02139, USA.

[14]Panjab University, Chandigarh 160 014, India.

[15]KLTE-ATOMKI, H-4010 Debrecen, Hungary[¶].

[16]UCD School of Physics, University College Dublin, Belfield, Dublin 4, Ireland.

[17]INFN Sezione di Firenze and University of Florence, I-50125 Florence, Italy.

[18]European Laboratory for Particle Physics, CERN, CH-1211 Geneva 23, Switzerland.

[19]World Laboratory, FBLJA Project, CH-1211 Geneva 23, Switzerland.

[20]University of Geneva, CH-1211 Geneva 4, Switzerland.

[21]University of Hamburg, D-22761 Hamburg, Germany.

[22]Chinese University of Science and Technology, USTC, Hefei, Anhui 230 029, China△.

[23]University of Lausanne, CH-1015 Lausanne, Switzerland.

[24]Institut de Physique Nucléaire de Lyon, IN2P3-CNRS, Université Claude Bernard, F-69622 Villeurbanne, France.

[25]Centro de Investigaciones Energéticas, Medioambientales y Tecnológicas, CIEMAT, E-28040 Madrid, Spain.

[26]Florida Institute of Technology, Melbourne, FL 32901, USA.

[27]INFN-Sezione di Milano, I-20133 Milan, Italy.

[28]Institute of Theoretical and Experimental Physics, ITEP, Moscow, Russia.

[29]INFN-Sezione di Napoli and University of Naples, I-80125 Naples, Italy.

[30]Department of Physics, University of Cyprus, Nicosia, Cyprus.

[31]Radboud University and NIKHEF, NL-6525 ED Nijmegen, The Netherlands.

[32]California Institute of Technology, Pasadena, CA 91125, USA.

[33]INFN-Sezione di Perugia and Università Degli Studi di Perugia, I-06100 Perugia, Italy .

[34]Nuclear Physics Institute, St. Petersburg, Russia.

[35]Carnegie Mellon University, Pittsburgh, PA 15213, USA.

[36]INFN-Sezione di Napoli and University of Potenza, I-85100 Potenza, Italy.

[37]Princeton University, Princeton, NJ 08544, USA.

[38]University of Californa, Riverside, CA 92521, USA.

[39]INFN-Sezione di Roma and University of Rome, "La Sapienza", I-00185 Rome, Italy.

[40]University and INFN, Salerno, I-84100 Salerno, Italy.

[41]University of California, San Diego, CA 92093, USA.

[42]Bulgarian Academy of Sciences, Central Lab. of Mechatronics and Instrumentation, BU-1113 Sofia, Bulgaria.

[43]The Center for High Energy Physics, Kyungpook National University, 702-701 Taegu, Republic of Korea.

[44]National Central University, Chung-Li, Taiwan, China.

[45]Department of Physics, National Tsing Hua University, Taiwan, China.

[46]Purdue University, West Lafayette, IN 47907, USA.

[47]Paul Scherrer Institut, PSI, CH-5232 Villigen, Switzerland.

[48]DESY, D-15738 Zeuthen, Germany.

[49]Eidgenössische Technische Hochschule, ETH Zürich, CH-8093 Zürich, Switzerland

## Note

[§]Supported by the German Bundesministerium für Bildung, Wissenschaft, Forschung und Technologie.

[‡]Supported by the Hungarian OTKA fund under contract numbers T019181, F023259 and T037350.

[¶]Also supported by the Hungarian OTKA fund under contract number T026178.

Supported also by the Comisión Interministerial de Ciencia y Tecnologia.

Also supported by CONICET and Universidad Nacional de La Plata, CC 67, 1900 La Plata, Argentina.

△Supported by the National Natural Science Foundation of China.

‡Deceased.